\documentclass[useAMS,usenatbib]{mn2e}
\usepackage{lscape}%
\usepackage{epstopdf}%
\usepackage{epsfig}%
\usepackage{natbib}
\usepackage{graphics}
\usepackage{longtable}

\title[Formation of cD Galaxies and their Parent Clusters]
{On the Formation of cD Galaxies and their Parent Clusters}
\author[Hrant~M.~Tovmassian, Heinz Andernach] {Hrant~M.~Tovmassian$^{1}$\thanks{E-mail:
hrant@inaoep.mx (HMT); heinz@astro.ugto.mx (HA)} and Heinz Andernach$^{2\star}$  \\
$^{1}$Instituto Nacional de Astrof\'{\i}sica, \'Optica y Electr\'onica,
AP 51 y 216, 72000, Puebla, Pue, Mexico \\
$^{2}$Departamento de Astronom{\'{i}}a, Universidad de Guanajuato, Apartado
Postal 144, 36000 Guanajuato, Gto, Mexico}

\begin{document}
\maketitle


\begin{abstract}
In order to study the mechanism of formation of cD~galaxies we search
for possible dependencies between the $K$-band luminosity of cDs and the
parameters of their host clusters which we select to have a dominant
cD~galaxy, corresponding to a cluster morphology of Bautz-Morgan (BM)
type~I.  As a comparison sample we use cD~galaxies in clusters where
they are not dominant, which we define here as non-BM\,I (NBMI) type clusters.
We find that for 71~BM\,I clusters the absolute $K$-band luminosity of
cDs depends on the cluster richness, but less strongly on the cluster
velocity dispersion. Meanwhile, for 35 NBMI clusters the correlation
between cD~luminosity and cluster richness is weaker, and is absent between
cD~luminosity and velocity dispersion. In addition, we find that the
luminosity of the cD~galaxy hosted in BM\,I clusters tends to increase
with the cD's peculiar velocity with respect to the cluster mean velocity.
In contrast, for NBMI clusters the cD~luminosity decreases with 
increasing peculiar velocity. Also, the X-ray luminosity of BM\,I
clusters depends on the cluster velocity dispersion, while in NBMI
clusters such a correlation is absent.  These findings favour the
cannibalism scenario for the formation of cD~galaxies. We suggest that cDs
in clusters of BM\,I type were formed and evolved preferentially in one and
the same cluster. In contrast, cDs in NBMI type clusters were either 
originally formed in clusters that later merged with groups or clusters 
to form the current cluster, or are now in the process of merging.

\end{abstract}


\section{Introduction}

The formation mechanism of the brightest cluster galaxies (BCGs) is
an important problem of modern astronomy (e.g.\ Lin \& Mohr 2004; 
von\,der\,Linden et al.\ 2007; Hansen et al.\ 2009; Garijo, Athanassoula, \&
Garcia-G\'{o}mez 1997; Tutukov, Dryumov, \& Dryumova 2007; Jord\'{a}n
et al.\ 2004). Some of the BCGs are cD~galaxies (Matthews, Morgan, \& Schmidt
1964)  which are characterized by an extended ``envelope'' or halo. The physical
properties of these unique objects were reviewed e.g.\ by Tonry (1987), Kormendy
\& Djorgovski (1989), Schombert (1992)\footnote{cf.\
{\footnotesize ned.ipac.caltech.edu/level5/March07/Schombert/frames.html}}
and Jord\'{a}n et al.\ (2004).

According to one of the proposed scenarios, BCGs are formed in cluster
cooling flows, when the gas density has grown enough to cool and condense,
leading to star formation in the cluster core (Silk 1976; Cowie \&
Binney 1977; Fabian 1994). In this scenario there should be color
gradients of the optical haloes in the sense that the latter should
become redder with increasing radius. However, such gradients have not 
been found (Andreon et al.\ 1992). Also, the finding that the X-ray gas
does not cool significantly below a threshold temperature of $kT\approx
1-2$\,keV (Kaastra et al.\ 2001; Peterson et al.\ 2001; Tamura et al.\
2001) puts this possibility of cD formation in doubt.

The second hypothesis on the formation of cDs supposes a rapid merging of
galaxies during cluster collapse (e.g., Merritt 1983). However, as Merritt
(1985) argues, the truncation of galaxy haloes during cluster collapse
would lead to time scales for dynamical friction longer than a Hubble
time and thus ``turn off" subsequent evolution in the cluster, i.e.\ the
growth rates after the cluster's virialization are slowed down.  Also,
according to simulations made by Dubinski (1998), the central galaxy does
not develop the extended envelope that is characteristic of cD~galaxies.

cD~galaxies which formed by the above-mentioned scenarios are expected to be
located near to the centres of their host cluster and are expected to have
a radial velocity close to the mean of the cluster galaxies. Meanwhile,
some cDs are located at an appreciable projected distance from the geometric
centre of the cluster and their median absolute peculiar velocity with 
respect to their host cluster's mean velocity is $\sim$27 per cent of the host
cluster's velocity dispersion (e.g.\ Oegerle \& Hill 2001, Coziol et al.\
2009, and references therein). This fact poses problems for the mentioned
mechanisms of formation of BCGs.

The third hypothesis for the cD formation is {\it galactic cannibalism}
(Ostriker \& Hausman 1977; Searle, Sargent \& Bagnuolo 1973; Ostriker, \&
Tremaine, 1975; Hausman \& Ostriker 1978; White 1976; Dressler 1980;
Barnes 1989; Baier \& Schmidt 1992; Garijo et al.\ 1997).
It appears to be the one that is most compatible with observational evidence.
According to this mechanism, cDs are formed as a result of
galaxies falling in along primordial filaments and their subsequent merging
(e.g.\ West et al.\ 1995; Fuller et al.\ 1999; Garijo et al.\
1997; Dubinski 1998; Knebe et al.\ 2004; Torlina et al.\ 2007). The weak trend
of the optical major axis of the BCGs to be aligned with their parent clusters' major
axes (Binggeli 1982; Struble 1987; Rhee \& Katgert 1987; Lambas, Groth, \&
Peebles 1988) supports the hypothesis of the formation of the former as a
result of hierarchical merging (Niederste-Ostholt et al.\ 2010). Mergers of
red galaxies, apparently without significant merger-triggered star formation
(dry mergers), have been observed at low redshift (e.g.\ van Dokkum 2005). 
According to Arag\'on-Salamanca et al.\ (1998), Gao et al.\ (2004), 
De\,Lucia \& Blaizot (2007), the stellar mass of BCGs grows
by a factor of between 3 and 4 via mergers since $z=1$. On the other hand, it has
been argued (e.g.\ Merritt 1985; Tremaine 1990) that the observed
dominance of BCGs cannot be achieved via cannibalism of other cluster
members, since the high velocity dispersion of clusters makes frequent
merging of galaxies unlikely. By studying the surface brightness and color
profiles of a few cD~galaxies and analysis of their globular cluster systems
Jord\'an et al.\ (2004) concluded that cDs appear to have formed rapidly 
(e.g., Dubinski 1998) at early times, via hierarchical merging prior to cluster virialization.

A related mechanism for formation of cDs involves tidal stripping
by cluster galaxies which pass near the cluster centre. The stripped material
falls to the centre of the potential well and may form the halo of the giant
galaxy there (Gallagher \& Ostriker 1972; Richstone 1975, 1976). Garijo et al.\
(1997) mention that this theory cannot
explain, however, the difference between central dominant cluster galaxies
with and without a prominent halo, and that the velocity dispersion
of stars in cD haloes is three times smaller than the velocity dispersion of
galaxies in the cluster. So this theory has a difficulty in explaining why
the tidally stripped material is slowed down as it builds up a cD halo.

In this paper we present arguments in favour of the cannibalism model
of the formation of cD~galaxies. We look for correlations between the 
cD~luminosity and its host cluster parameters, including the number of 
its members, which was not considered in other models. Our emphasis is on the formation
of cD~galaxies in clusters of Bautz-Morgan (BM) type~I (Bautz \& Morgan
1970), since the performed analysis is applicable only to clusters with a
single dominant galaxy. For comparison we considered a sample of clusters
containing cD~galaxies as well as one or more other galaxies of comparable
luminosity, and call the latter clusters ``non-BM\,I type'' (or NBMI in what 
follows).  The observational data we used allows us to suggest that
clusters of BM\,I and NBM\,I types have different evolution histories.

\section{The Data}

In the analysis presented here we looked for possible correlations between
the $K$-band luminosity of cD~galaxies on the one hand, and the cluster
richness and the velocity dispersion on the other. For the selection of
clusters we started out from the compilation of BCGs in Abell clusters
(Abell et al.\ 1989) by Coziol et al.\ (2009), using clusters of any
BM~type containing a cD~galaxy, but restricting ourselves to clusters with
redshift $z<0.15$.  In compiling our list we excluded all supplementary
S-clusters, and exluded most clusters that had more than one significant
redshift components along the line of sight. We required that the
mean redshift be based on at least five spectroscopic members. However,
in the analysis involving the cluster velocity dispersion $\sigma_v$ we 
only used those clusters with at least 10 cluster member redshifts. 
We took the cluster velocity dispersions $\sigma_v$ from the most recent
version of the Abell cluster redshift compilation maintained by one of us
(see Andernach et al.\ 2005 for a description). A few velocity dispersions
were taken from a recent analysis by Zhang et al.\ (2011).

We used the Abell number count, $N_A$, as an indicator of the cluster
richness. $N_A$ is the number of galaxies in the magnitude range between
$m_3$ and $m_{3}+2$, where $m_3$ is the apparent photored magnitude of
the third-brightest cluster member, located within one Abell radius,
$R_A$, of the cluster centre, where $R_A = 1.7'/z$. The values of $N_A$
were taken from Abell et al.\ (1989) and were mostly based on estimated
redshifts used to determine the Abell radii. Thus we understand that $N_A$
is not a precise measure of the cluster richness. Nevertheless, it is
an appropriate parameter, since it gives the number of galaxies in the
central region of a cluster where merging of galaxies preferentially
takes place.  For those clusters which had other significant components
at different redshift along the line of sight, we corrected the Abell count 
$N_A$ downwards, in proportion to the number of 
measured redshifts in the component containing the cD~galaxy, as compared
to the number of redshifts in all components along the line of sight
(see the values marked with an asterisk in column~6 of Table~1 below).

\subsection{Definition of main and control sample}

In our study we considered separately the BM\,I type clusters with a dominant
cD~galaxy and NBMI clusters. According to Bautz \& Morgan (1970) the 
BM\,I clusters are defined as {\it clusters containing a centrally 
located cD galaxy}. In BM\,II types the brightest galaxies 
{\it are intermediate in appearance between class cD and the Virgo-type 
giant ellipticals}. BM\,III types were defined as {\it clusters containing 
no dominant galaxies}. We introduced a quanitative criterion to differentiate
between clusters. We assumed a cD~galaxy as dominant and the cluster as of 
BM\,I type, if the cD's $K$-band magnitude was brighter than the 
second-brightest cluster member by  $\Delta K\ge1.00^m$. 
This corresponds to a luminosity of the brightest galaxy 2.5~times 
higher than that of the second-brightest galaxy. When this ``luminosity gap''
was less than a factor of 1.9, i.e.\ the $K$-band magnitude difference was
less than $0.70^m$, we assumed that the cluster was of NBMI type. To avoid
the ambiguity of finding the second-brightest galaxy in a cluster we 
imposed a lower limit of 0.035 for the cluster redshift. Since the clusters 
with a luminosity gap between the first and second-brightest galaxy in the 
range $0.7^m<\Delta K<1.0^m$ may belong to either of the BM\,I or 
NBMI classes, we omitted these intermediate clusters. We list the
luminosity gap $\Delta K$ between the cD and 2nd-brightest cluster
member in column~4 of Table~1.

For the determination of the cD~galaxy luminosity we used the $K_{s-total}$
apparent magnitude from the 2MASS Extended Source Catalogue (Jarrett
et al.\ 2000). The 2MASS magnitudes have been widely used in galaxy
studies (e.g.\ Temi, Brighenti, \& Mathews 2008; Courteau et al.\ 2007;
Masters, Springob, \& Huchra, 2008, etc.). The $K$~band is more appropriate
for our study, since it encompasses
the light of the predominantly red population in early-type galaxies. Note
that Lauer et al.\ (2007) showed that 2MASS photometry is free from possible
errors which may be caused by the sky background subtraction and
crowding. The most important inconsistency may be produced by the
extrapolation scheme to generate ``total magnitudes" (Jarrett et al., 2000).
Lin \& Mohr (2004) used a correction scheme to extrapolate isophotal
magnitudes to ``total" magnitudes and showed that both schemes are
consistent. Bell et al.\ (2003) mentioned that 2MASS magnitudes have
problems in detecting the low surface brightness light, such as haloes of
cD~galaxies (e.g.\ Schombert 1988). In addition, Lauer et al.\ (2007)
demonstrated that 2MASS photometry is likely to underestimate the total
light from cDs. However, it is obvious that the errors in the 2MASS
$K$-magnitudes may not create correlations of absolute magnitude $M_K$
with the corresponding cluster parameters, $N_A$ and $\sigma_v$, which
also are determined with some errors. The errors may only increase
the dispersion and thus dilute or even destroy the correlations we are
seeking.

The absolute stellar magnitudes $M_K$ of the cD~galaxies were deduced
using the mean redshift of their host cluster (Andernach et al.\ 2005),
adopting a Hubble constant of $H_0=72$\,km\,s$^{-1}$\,Mpc$^{-1}$.
A correction for the Galactic extinction was introduced according to
Schlegel, Finkbeiner \& Davis (1998) as given in the NASA/IPAC Extragalactic
Databae (NED, ned.ipac.caltech.edu),
and the k-correction according to Kochanek et al.\ (2001).

\subsection{Lists of BM\,I and NBMI type clusters}

Based on the BCG compilation by Coziol et al.\ (2009), we inspected images of 
all clusters containing a BCG classified as a cD~galaxy,
and compared $K_{s-total}$ magnitudes of the brightest and second-brightest 
galaxies using NED's photometric data. During this inspection we found
that a few clusters listed as of BM\,I type in Coziol et al.\ (2009) are in
fact clusters of NBMI type according to our definition above. 
For example, in the supposed BM\,I type cluster A1839 the $K$~magnitudes of the 
brightest (2MASX J14023276$-$0451249) and the second-brightest
(2MASX J14023417$-$0449449) galaxies are about the same: $12.84^m$ and 
$12.81^m$. 
We also found examples of the opposite case: the second-brightest galaxy 
(2MASX J14070976+0520132) in the supposed BM\,II type cluster A1864A is fainter 
than the cD (2MASX J14080526+0525030) by 1.15$^m$, so we consider the cluster 
as of BM\,I type.

In addition to the Coziol et al.\ (2009) sample of clusters we made use
of an additional set of galaxies claimed to be cD~galaxies in NED, kindly 
provided to one of us (H.A.) by H.G.~Corwin~Jr. in 2006. We inspected Digitized
Sky Survey images of those cDs that are located within Abell clusters of
a sufficient number of measured redshifts.  As a result we compiled a
list of 71~cDs in clusters of BM\,I type and of 35~cDs in clusters of NBMI type,
presented in Table~1. The 22 intermediate-type clusters out of 128 listed 
in Table~1 were omitted from the analysis.

\section{Results}

In this section we discuss the correlations between six different
pairs of parameters we collected for our cluster sample. \\[-.5ex]

\noindent
(a)~ The distribution of absolute $K$-magnitudes of cD~galaxies is 
shown in Figure~1, separately for BM\,I and NBMI clusters, versus the 
redshift of their host clusters. The luminosity of the most luminous
cDs increases gradually with increasing redshift~$z$, forming an upper
envelope of the $z - M_K$ distribution. Less luminous cD~galaxies are 
observed almost equally all over the considered redshift range. \\[-.5ex]

\begin{figure}
\centering
\psfig{file=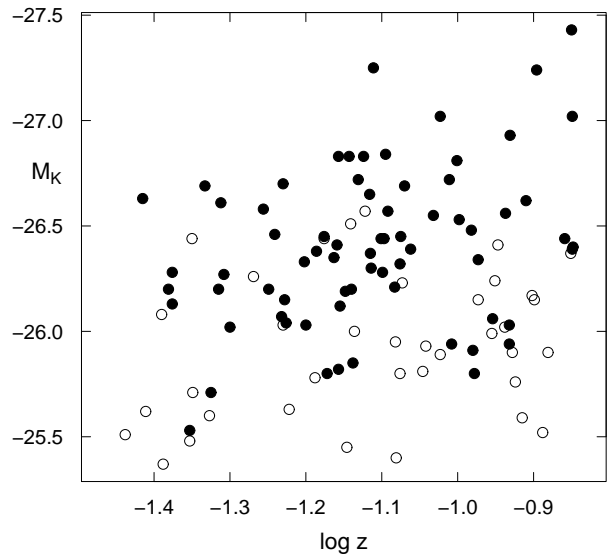,width=8.cm,angle=0,bbllx=1pt,bblly=9pt,bburx=478pt,bbury=456pt,clip=}
\caption{$M_K$ absolute magnitude of cD~galaxies versus redshift $z$
in clusters of BM\,I type (filled circles) and NBMI type (open circles).}
\end{figure}

\noindent
(b)~ In Figure~2 we present the graphs of $log N_A$ vs.\ $log~z$
separately for clusters of type BM\,I and NBMI. Figure~2 shows that
the Abell number count $N_A$ of both BM\,I and NBMI clusters is weakly
rising with redshift. This reflects the well-known effect that at higher
redshifts the poor clusters are missed and the relative fraction of
rich clusters increases (Scott 1957; Postman et al.\ 1985). Over the
considered redshift range of 0.035 to 0.15 the average $N_A$ in BM\,I
clusters increases from from $N_A\approx43$ to $N_A\approx74$. In the
case of NBMI clusters $N_A$ increases from approximately 49 to 77. \\[-.5ex]

\begin{figure}
\psfig{file=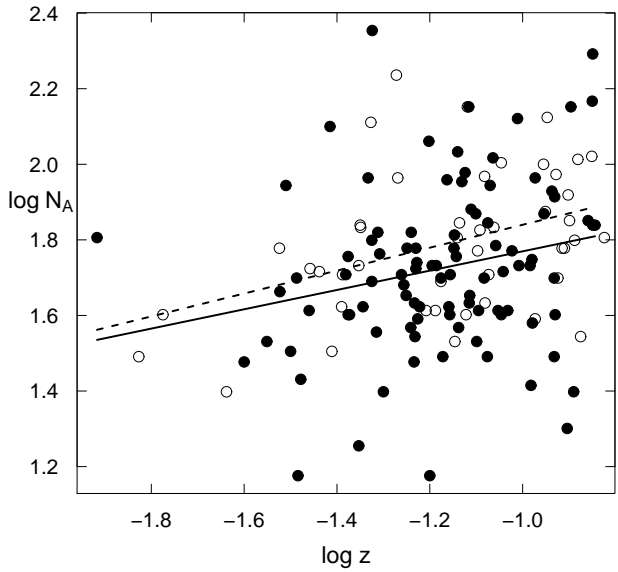,width=8.2cm,angle=0,bbllx=1pt,bblly=9pt,bburx=478pt,bbury=456pt,clip=}
\caption{The Abell number count $N_A$ of clusters of BM\,I and NBMI types
that host the cD~galaxy, versus the cluster redshift. Symbols are as in Fig.~1,
and continuous and dashed regression lines correspond to BM\,I and NBMI 
clusters, respectively.}
\end{figure}

\noindent
(c)~ Whiley et al.\ (2008) found a weak dependence of the cD~luminosity
on the velocity dispersion $\sigma_v$ of the cluster. We looked for a 
correlation between the luminosity of the cD~galaxy (expressed as $M_K$) and 
$\sigma_v^2$, since the mass $M$ of a cluster depends on the square of the
velocity dispersion of the parent cluster. Figure~3 shows that the $M_K$
magnitude of cD~galaxies in clusters of BM\,I type certainly correlates with
$\sigma_v^2$, with a correlation coefficient of $-0.50$, and a regression
slope of $-0.65\pm0.13$. Meanwhile, the absolute magnitude $M_K$ of
the cD~galaxies in NBMI clusters does not correlate with $\sigma_v^2$
of the cluster. The correlation coefficient is $-0.11$. \\[-.5ex]

\begin{figure}
\psfig{file=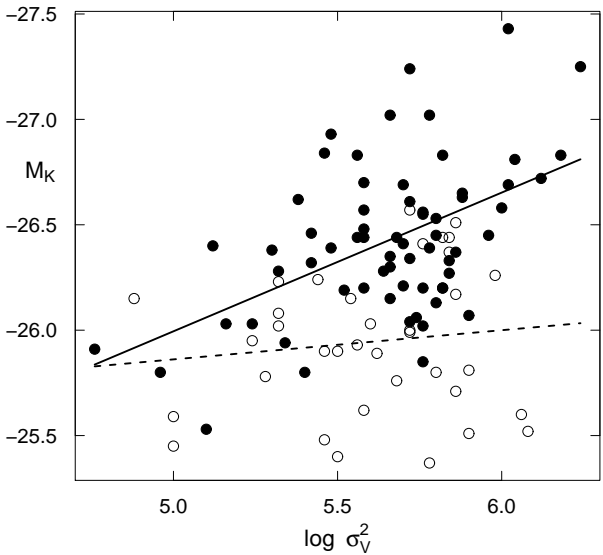,width=8.cm,angle=0,bbllx=1pt,bblly=9pt,bburx=478pt,bbury=456pt,clip=}
\caption{The absolute $K$-magnitude of cD~galaxies versus the square
of the velocity dispersion $\sigma_v^2$ of the parent clusters of BM\,I 
and NBMI types. Symbols are as in Fig.~1, and regression lines as
in Fig.~2.}
\end{figure}

\noindent
(d)~ A weak correlation between the BCG luminosity and cluster 
richness has been found previously (e.g.\ Schneider, Gunn, \& Hoessel 1983; 
Schombert 1987). In Figure~4 we plot the absolute $M_K$-magnitude of 
cD~galaxies versus the corresponding $N_A$ of their host clusters separately
for BM\,I and NBMI types. Figure~4 shows that the $K$-band luminosity of cDs in
clusters of BM\,I type correlates with $N_A$. The correlation coefficient is
$-0.62$, and the slope of the regression line is $-1.09\pm0.17$. Meanwhile,
the luminosity of cDs in NBMI clusters shows a weaker dependence on the
cluster richness, with a correlation coefficient of $-0.21$ and a slope of
$-0.43\pm0.34$. \\[-.5ex]

\begin{figure}
\psfig{file=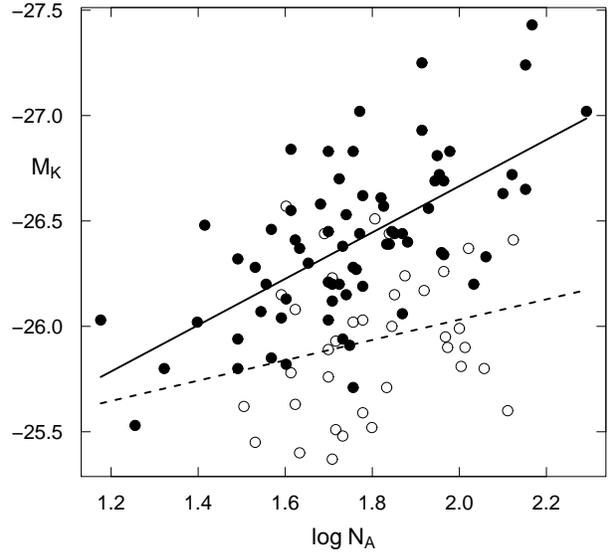,width=8.cm,angle=0,bbllx=1pt,bblly=9pt,bburx=478pt,bbury=456pt,clip=}
\caption{$M_K$ absolute magnitude of cD~galaxies versus the Abell
number count, $N_A$, of clusters of BM\,I and NBMI types that host a cD~galaxy. 
Symbols are as in Fig.~1, and regression lines as in Fig.~2.}
\end{figure}

\noindent
(e)~ One may expect that the velocity dispersion of a cluster would
depend on its richness. In Figure~5 we present the graph $log N_A$ vs.\ 
$log \sigma_v$ separately for clusters of BM\,I and NBMI types. It shows that
the velocity dispersion of both BM\,I and NBMI clusters increases with
increassing cluster richness. The correlation coefficients are about the
same, 0.43 and 0.46, respectively. The slopes are different: $0.28\pm0.06$
for clusters of BM\,I type and steeper, $0.41\pm0.13$ for NBMIs. \\[-.5ex]

\begin{figure}
\psfig{file=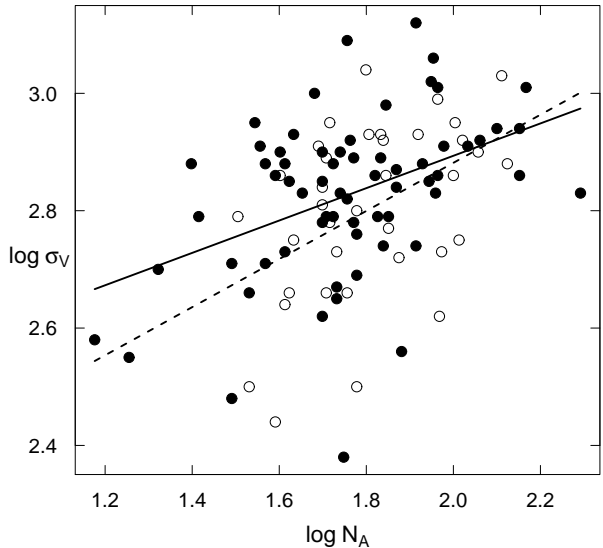,width=8.cm,angle=0,bbllx=1pt,bblly=9pt,bburx=478pt,bbury=456pt,clip=}
\caption{The cluster velocity dispersion, $\sigma_V$, versus the 
Abell number count $N_A$ for clusters of BM\,I and NBMI types. 
Symbols are as in Fig.~1, and regression lines as in Fig.~2.}
\end{figure}

\noindent
(f)~ It has been found that some cD~galaxies have peculiar velocities,
defined as the difference between the BCG and the cluster mean radial
velocity: ~ $v_{pec} = (v_{BCG} - cz_{cl})/(1 + z_{cl})$. In some clusters
these peculiar velocities may reach significant fractions of the cluster
velocity dispersion (Sharples, Ellis, \& Gray 1988; Hill et al. 1988;
Oegerle \& Hill 1994; Pimbblet, Roseboom, \& Doyle 2006; Coziol et
al. 2009). We tried to find out whether the cD~galaxy luminosity depends
on its peculiar velocity. In Figure~6 we plot $M_K$ vs.\ $log\,|v_{pec}|$
for cDs in clusters of BM\,I and NBMI types. Figure~6 shows that the $K$-band
luminosity of cD~galaxies in BM\,I clusters increases with $v_{pec}$,
but shows the opposite trend in NBMI clusters. We omitted the cluster
A2657 from this plot because for its very low $v_{pec}\approx0$, placing it far
from the bulk of the other clusters. The correlation coefficients for both
samples are low, $-0.25$ and 0.49, respectively, while the slopes of
the regression lines, $-0.21\pm0.11$ and $0.43\pm0.14$ respectively,
differ significantly from each other.  \\[-.5ex]

Note that the correlations found in the above items (a) to (f) are
revealed in spite of possible errors in the used parameters of clusters.
Obviously, the errors may only weaken any existing correlations.

We wish to note also that for Figs.~2--6 we used the ``robust fitting of
linear models'' ({\tt rlm} in the R software package), and in all cases the robust
fit was indistinguishable from the standard linear model ({\tt lm}), i.e.\ it
differed much less than the errors of the fit parameters. In what follows
we shall discuss seven aspects which we find to support our conclusions.

\begin{figure}
\psfig{file=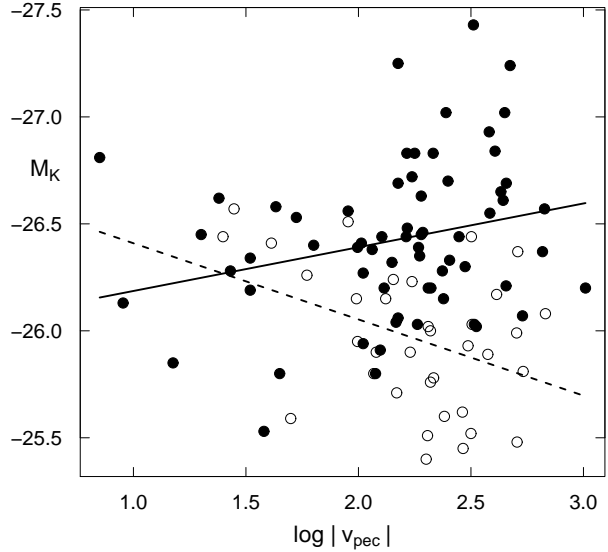,width=8.cm,angle=0,bbllx=1pt,bblly=9pt,bburx=478pt,bbury=456pt,clip=}
\caption{cD~luminosity $M_K$ vs.\ the peculiar velocity for cD~galaxies in
clusters of BM\,I and NBMI types. Symbols are as in Fig.~1, and
regression lines as in Fig.~2.}
\end{figure}

\section{Discussion and Conclusions}

As mentioned above, we define a BM\,I type cluster as one with a single
dominant cD~galaxy, and a NBMI type cluster as one that contains one or
more galaxies with luminosities comparable to that of the cD~galaxy.
In this section we argue that the correlations we found between the
parameters of cD~galaxies and their parent clusters not only reveal
differences in the formation histories between BM\,I and NBMI type
clusters, but also favour the cannibalism model for the cD galaxy
formation. \\[-1ex]

\noindent
{\it 1. Different evolution histories of BM\,I and non-BM\,I clusters 
hosting a cD galaxy} \\

According to hierarchical model, clusters evolve by merging with groups
of galaxies and other clusters (e.g.\ Merritt, 1984; Zabludoff \& Mulchaey,
1998). We found that clusters of BM\,I and NBMI types have different
properties that give clues to their different evolution histories.

The dependence of the $K$-band luminosity of cD~galaxies on the host
cluster richness expressed by the Abell number count $N_A$  is stronger
in BM\,I clusters (cf.\ Fig.~4). The slope of the regression line of the
correlation $N_A-M_K$ in BM\,I clusters is $-1.10$, while in
NBMI clusters the slope is only $-0.48$. Also the cD~galaxy luminosity 
hosted in BM\,I clusters depends on the cluster velocity dispersion, 
$\sigma_v^2$ (Fig.~3). The correlation coefficient is
$-0.50$, and the slope of the regression line is $-0.65$. Meanwhile,
the cD~luminosity in NBMI clusters does not depend on the host cluster
velocity dispersion.

The correlations of the cD~luminosity with the parent cluster parameters
for BM\,I clusters allow us to suggest that cD~galaxies in these
clusters were formed and evolved preferentially within one and the same
cluster. The absence of the cD~luminosity correlations with the parent
cluster parameters for NBMI type clusters shows that the cD~galaxy in
them was formed in a cluster that is in the process of merging with
another cluster, or has already merged with other groups or clusters. 
The parameters of the composite cluster will obviously differ from those
of the {\it initial} cluster in which the cD~galaxy was formed, and
correlations observed in BM\,I clusters will be weakened or erased in
composite NBMI clusters. Also, the velocity dispersion of the composite
cluster will not be proportional to the cluster mass.

We suppose that the luminosity of the cD~galaxy formed in the initial
cluster would fit the correlations seen in Figures~3 an 4. Merging of other
groups and clusters with the initial cluster will increase the richness
and velocity dispersion of the resulting observed cluster, while the
luminosity of the cD~galaxy will remain the same. The richer and more
massive the initial cluster is, (and consequently the brighter the
formed cD~galaxy), the more groups will merge with it, the larger will
be the increase of richness and velocity dispersion, and the farther 
to the right from the regression line determined by BM\,I clusters 
in Figures~3 and 4 the corresponding point will be located. As a result, 
the slope of the regression line of $M_K$ vs.\ $N_A$ for NBMI clusters will
decrease. Since the correlation between $M_K$ and $\sigma_v^2$ for BM\,I 
clusters is generally weaker, the correlation for NBMI clusters even 
disappears.

The steeper slope of the regression line for NBMI clusters in Figure~5
also favours the suggestion made on the different evolution histories of BM\,I
and NBMI clusters. Merging of groups and clusters with the initial cluster
will increase the richness and velocity dispersion of the observed
cluster. If the mean velocity of member galaxies of merged groups
differs significantly from that of the initial cluster, the increase of
the velocity dispersion will obviously be stronger than the increase in
richness. Therefore, the slope of the regression line for NBMI clusters
in the graph $N_A-\sigma_v$ becomes steeper than for BM\,I type clusters.

The cluster(s) that merge with the initial cluster (forming the cD) 
are generally poorer, and their brightest galaxy will usually be fainter
than the cD~galaxy in the initial cluster. However, it is possible that
the luminosity of the BCG in the cluster that merges with the initial one
is comparable to, or even brighter than that of the cD~galaxy. This may
be the case for some NBMI clusters in our sample (A1736B, A2051, A2969) 
where the cD~galaxy is even fainter than the brightest galaxy
formed in the merged (currently seen) cluster.

If the clusters of BM\,I and NBMI types indeed have different evolution
histories, and the velocity dispersion of NBMI clusters is not proportional
to cluster mass, then one may expect that the X-ray properties of
both types of clusters will be different. In order to check this
conjecture we compared the dependence of X-ray luminosity on the cluster
velocity dispersion for both types of clusters. In Figure~7 we plot the
cluster X-ray luminosity, $log\,L_{X,500}$, versus $\sigma_v$ for 
BM\,I and NBMI clusters. $L_{X,500}$ is the luminosity within $r_{500}$, 
the radius within which the mean overdensity of the cluster is 
500~times the critical density of the Universe at the cluster redshift, as
published by Piffaretti et al.\ (2011). Figure~7 shows that, as we 
expected, the X-ray luminosity of NBMI clusters does not
depend on the velocity dispersion, while in BM\,I clusters it does.
The correlation coefficient is 0.45, and the slope of the regression 
line, i.e.\ the power in the $L_{X,500} \propto \sigma_v^x$
is $x=1.60\pm0.57$.

\begin{figure}
\psfig{file=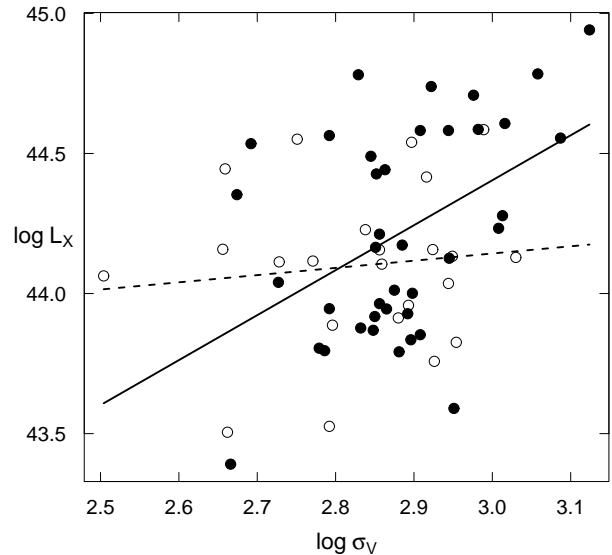,width=8.cm,angle=0,bbllx=1pt,bblly=9pt,bburx=478pt,bbury=456pt,clip=}
\caption{The cluster X-ray luminosity (within $r_{500}$ and in units
of erg\,s$^{-1}$) versus the velocity dispersion $\sigma_v$
for clusters of BM\,I and NBMI types. Symbols are as in Fig.~1, and
regression lines as in Fig.~2.}
\end{figure}

We conclude that clusters of BM\,I and NBMI types have different evolution 
histories. \\[-1ex]

\noindent
{\it 2. Luminosity difference between cD~galaxies in BM\,I and NBMI clusters} \\

One may expect that cD~galaxies formed and evolved in a single rich
cluster may be brighter than those cDs that were formed in relatively
poor clusters that later merged with other galaxy groups or clusters. 
Indeed, Figure~1 shows that cD~galaxies in BM\,I clusters are more luminous 
than those formed in NBMI clusters: the mean $M_K$ of cDs in clusters of 
BM\,I type is $-26.39\pm0.38$ (95 per cent confidence), while that of cDs in
NBMI clusters, $-25.96\pm0.34$ (95 per cent confidence), is fainter. The Kolmogorov-Smirnov (KS)
and Mann-Whitney~U (MWU) two-tailed tests show that the two samples of 
$M_K$ magnitudes are significantly different ($P_{KS}$ = 0.00001 and 
$P_{MWU}<0.0001$). If we restrict both samples to the most luminous cDs with 
$M_K<-26.0$, then the mean $M_K$ of cDs in BM\,I clusters will be
$-26.48\pm0.32$, and in NBMIs $-26.25\pm0.19$ (both 95 per cent confidence), 
i.e.\ there is still a luminosity difference in this restricted sample. 
Hence, different luminosities of cD~galaxies in BM\,I and NBMI clusters favour 
the suggestion of different evolution histories of the two types of clusters. \\[-1ex]

\noindent
{\it 3. Dependence of the cD~galaxy luminosity on its peculiar velocity} \\

Figure~6 shows that the luminosity of the cD~galaxy in clusters of BM\,I
type increases with the peculiar velocity of the cD~galaxy, while it
shows the opposite trend in clusters of NBMI clusters. A different 
dependence of the cD~luminosity on its peculiar velocity in clusters 
of BM\,I and NBMI clusters is explained within the assumed model of a
different evolution of BM\,I and NBMI clusters.

We suggest that the increase of the cD~galaxy luminosity hosted  
in BM\,I clusters may be explained in the following way. The cD~galaxy may
be formed not at the exact gravitational centre of the corresponding
cluster and will oscillate about it (Quintana \& Lawrie, 1982). The
higher the velocity of its movement, the more chances it will have to
encounter with other members of the cluster and the larger will be the
number of galaxy mergers. Therefore, cDs with higher peculiar velocity
become more luminous. This fact favours the cannibalism model of the cD
galaxy formation.

We found that the cD~luminosity hosted in BM\,I clusters depends on the
cluster richness (cf.\ Fig.~4). Obviously the same must be the case in
the initial cluster that later became of NBMI type after its merging with
other groups or clusters. The poorer the initial cluster, the fainter will
be the formed cD~galaxy. At the same time, the richer the merged cluster,
the higher will be the peculiar velocity of the cD~galaxy, since the mean
redshift of the merged cluster will differ more from the redshift of the
initial cluster. Therefore, the fainter the cD~luminosity, the higher may
be its peculiar velocity, which is consistent with Fig.~6. This supports
the conclusion we made on the different evolution histories of clusters
of BM\,I and NBMI types. WE note also that cD~galaxies in NBMI clusters
may be located far from the bottom of the gravitational well of the
cluster. The projected separation of the BCG from the X-ray peak of the
corresponding cluster was measured by Hudson et al.\ (2010). Among the
brightest galaxies with a large separation there are clusters common to
our list: A0754, A1736 and A3376 with separations of 714, 642 and 939~kpc,
respectively. The first two are of NBMI type, and the third one, A3376,
by its $\Delta\,K=0.73$ value is close to an NBMI cluster, but for the
sake of reliability, we excluded this cluster from our analysis. Hence,
this fact also supports the suggested hypothesis on different evolution
of BM\,I and NBMI clusters. \\[-1ex]

\noindent
{\it 4. The difference of merging efficiency on cluster richness and velocity
dispersion} \\

According to all models of cD~galaxy formation, the cD~luminosity 
depends on the parent cluster mass. This is the case when the cD
was formed in the cluster cooling flow or by a rapid merging of galaxies
during cluster collapse. According to the cannibalism model, the higher 
the cluster mass, the stronger will be the gravitational force towards 
the cluster center, and consequently more member galaxies will be 
attracted to the central area and may be cannibalized.

The mass of a cluster may be estimated from its velocity dispersion. 
The cluster mass may be estimated also from the number of its member 
galaxies, characterized by the Abell number count $N_A$. Both 
parameters, $\sigma_v$ and $N_A$, are correlated (cf.\ Fig.~5).
Below we show that the cluster richness is more 
decisive for the formation of a cD~galaxy.

The luminosity of the cD~galaxy depends on the richness $N_A$ of the host
cluster. According to the regression line in Figure~4, an increase of 
$M_K$ by one magnitude from $-25.8$ to $-26.8$ in BM\,I clusters corresponds
to an increase of $N_A$ from 16 to 138. According to the regression line in
Figure~5, an increase of $N_A$ from 16 to 138 corresponds to an increase
of $\sigma_v$ from 468~km\,s$^{-1}$ to 840~km\,s$^{-1}$, i.e.\ by 1.8~times. 
If the effectiveness of the cluster richness and the velocity
dispersion for the formation of the cD~galaxy are about the same, we may
expect that for an increase of the luminosity from $-25.8$ to $-26.8$
the velocity dispersion must increase by about 1.8~times. However, the
regression line in Figure~3 shows that for such an increase of $M_K$
the velocity dispersion increases in fact from 190~km\,s$^{-1}$ to 
1340~km\,s$^{-1}$, i.e.\ by 7.1~times. This means that the cluster 
richness is by $7.1/1.8\approx4$ times more effective for the 
cD~formation than the velocity dispersion which is commonly used 
to estimate the cluster mass.

This finding strongly favours the cannibalism model of the cD~galaxy
formation. The higher the BM\,I cluster richness, the more of its members
may be cannibalized, and the more luminous will be the resulting 
cD~galaxy. Hence, the cluster richness, i.e.\ the number of member galaxies,
plays a more decisive role in the cD~formation than the cluster mass
determined by velocity dispersion. Models for cD~formation other than the 
cannibalism scenario do not differentiate between the mass of the 
cluster and the number of its members.

It is evident that in NBMI clusters the luminosity of the cD~galaxy
formed in the initial cluster may not depend on the parameters of the
(presently) observed cluster formed as a result of merging of the 
initial cluster with other galaxy groups or clusters. \\[-1ex]

\noindent
{\it 5. Rough estimate of the number of galaxies merged to form the cD} \\

Suggesting that cD~galaxies were assembled through the so-called
dissipationless ``dry'' mergers of gas-poor, bulge-dominated systems 
(Tran et al.\ 2005; van Dokkum 2005; Bell et al.\ 2006; De Lucia et al.\ 2006;
Bernardi et al.\ 2007) we roughly estimate the number $N_m$ of merged
galaxies required to form the observed cD. Dry mergers are consistent
with the high central densities of ellipticals and their old stellar
populations. If mergers are responsible for the formation of BCGs,
then, as has been shown by several authors (Malumuth \& Richstone 1984
and references therein), the luminosity growth will be at the expense
of fainter members. Assuming that the merged galaxies are ordinary
faint galaxies, with an absolute magnitude $M_{K(isol)} = -22.7$ (see
Tovmassian, Plionis \& Andernach 2004) for isolated E/S0 galaxies,
we estimate that the faintest cD with $M_K \approx -25.5$ are formed
by the assembly of only about 13 galaxies with a mean luminosity of an
isolated E/S0 galaxy. The most luminous cDs with $M_K \approx -27.5$
are formed by merging of about 80 ordinary E/S0 galaxies. \\[-1ex]

\noindent
{\it 6. The rate of galaxy merging in poor and rich clusters} \\

Merritt (1985) and Tremaine (1990) showed that the efficiency of merging
depends on the cluster velocity dispersion $\sigma_v$ in the sense that
a high velocity dispersion will prevent frequent merging. In agreement
with this, Forman \& Jones (1982), and Schombert (1987) mentioned that
a cluster with a lower velocity dispersion would have a higher rate
of mergers.

We compared the velocity dispersion $\sigma_v$ of clusters of BM\,I type
located near to the lower and upper envelopes of the $M_K-z$
distribution in Figure~1. The mean $\sigma_v$ for 7~clusters with the
least luminous cDs and known $\sigma_v$ (A0912A, A1076, A1227A, A2110,
A2170B, A2544, A3104) is $504\pm176$\,km\,s$^{-1}$. 
These clusters are poor. Their mean $N_A$ is $41\pm18$.
The mean $\sigma_v$ for 12 clusters with the most luminous cDs in Figure~1
(A0085A, A0399,  A0655, A0690A, A1146, A1644, A1738,  A2420, A2457,
A3112B, A3571, A4059) is $819\pm215$\,km\,s$^{-1}$.
The mean $N_A$ of these clusters is $86\pm35$. The MWU two-tailed tests
show that the difference of the velocity dispersion  $\sigma_v$ of
these two subsamples is highly significant $P_{MWU}$ = 0.0128). Hence,
the process of merging is fast in poor clusters of BM\,I type with small
velocity dispersion. The situation is different in clusters of NBMI
type, where no difference of $\sigma_v$ of clusters with the most and least
luminous cD~galaxies is observed. The mean $\sigma_v$ of the 12 NBMI clusters
hosting most luminous cD~galaxies (median $M_K=-26.3$) is $706\pm195$\,km\,s$^{-1}$ 
and that of the 9 clusters with the least luminous cD~galaxies 
(median $M_K=-25.5$) is $688\pm276$\,km\,s$^{-1}$. \\[-1ex]

\noindent
{\it 7. The evolution of cD~galaxies in rich and poor clusters} \\

The evolution of cD~galaxies in BM\,I clusters may be followed in the frames
of the adopted cannibalism model in Figure~1. Our Figure~2 suggests that
clusters at low redshift are poorer on average. However, part if this
can be explained by the fact that rich clusters are rare and the local 
volume is small.  Since the velocity dispersion of poor
clusters is small, the process of merging in them is fast. At the same
time, the reservoir of galaxies for merging is also small. Therefore,
the process of luminosity increase in cD~galaxies in poor clusters
terminates in a relatively short time, and it may reach only a modest
luminosity. The final stage of poor clusters may be a fossil group
(Tovmassian 2010). Thus, cD~galaxies in nearby poor clusters almost
reached their possible maximum, rather low luminosity.

The mean $N_A$ of the five poorest BM\,I clusters with $z<0.05$ 
(A0912A, A1308A, A2271, A0376, A1890) is $30\pm8$, and that of the
five poorest clusters in our highest distance range ($0.11<z<0.15$; 
A0038, A1023, A1068, A1076, A3854A) is $56\pm15$. Thus, the distant
poor clusters are relatively rich, the process of cannibalism in them
probably still continues, and cDs in these clusters did not yet reach
their possible maximum luminosity.

The situation is different in rich clusters. The velocity dispersion of
rich clusters is high. Therefore, the  rate of merging in rich clusters
is low and lasts longer also due to the larger reservoir of candidate
galaxies for merging. The cD~galaxies in rich clusters slowly move up in
Figure~1, occupying almost uniformly the space from the smallest to the
highest luminosities in each redshift range. Thus, the upper envelope of
this distribution may be explained without invoking the Malmquist bias,
but simply by the fact that cDs are observed in clusters of different
richness and assuming cannibalism for their formation

Hence, the observational data favour the cannibalism model of the cD
galaxy formation. We conclude that cD~galaxies in clusters of BM\,I type
were formed and evolved in one and the same cluster. We suggest also
that cDs in NBMI clusters were originally formed in poorer cluster and
are observed now in clusters that were formed by merging with other
galaxy groups and clusters. \\[-1ex]

\section*{Acknowledgements}

This research has made use of the NASA/IPAC Extragalactic Database
(NED) which is operated by the Jet Propolsion Laboratory, California
Institute of Technology, under contract with the National Aeronautics
and Space Administration. The figures were created within the ``R"
software package (cran.r-project.org) and we thank R.\,A.~Ortega Minakata
for advice in producing these. We thank the anonymous referee whose comments 
helped us to substantially improve the paper.


\onecolumn

\begin{table*}
\centering
\begin{minipage}{180mm}
\caption{Data on the 71 clusters of type BM\,I, 22 of intermediate type, and 35 of type NBMI
and corresponding cD~galaxies as well as second-brightest galaxies.
Table columns are as follows:~
(1) Abell cluster designation, appended by a letter indicating the cluster's component along the line of sight; 
(2) mean redshift of the cluster; 
(3) number of galaxies which were used to determine the mean cluster redshift and velocity dispersion; 
(4) difference in $K$-band magnitude between cD and 2nd-brightest galaxy; 
(5) absolute $K$-band magnitude, $M_K$, of the cD~galaxy; 
(6) Abell number count $N_A$; ``*" indicates a downward correction 
      to allow for an overlap of two or more redshift components of a cluster;
(7) cluster velocity dispersion $\sigma_v$;  
(8) peculiar velocity of cD~galaxy; 
(9) NED name of cD~galaxy;
(10) NED name of 2nd-brightest galaxy. [Correction added after online publication 2012 November
26: Second-brightest galaxy ID corrected for cluster A0690A.] Additional correction made in the present
astro-ph version on 1-Dec-2012: the 2nd-brightest galaxy in A1149 is actually 2MASX~J11032040+0730463 
at z=0.0714, implying $\Delta K=0.15$ for A1149.}
\begin{tabular}{@{}lcrrcrrrll@{}}
\hline
Abell & $z$ & $N_z$ & $\Delta K$ & $M_K$ & $N_A$ & $\sigma_v$ & $v_{pec}$ & \multicolumn{1}{c}{cD~galaxy ID} & \multicolumn{1}{c}{2nd-brightest~galaxy ID} \\
Cluster  &  &       &   mag     &  mag  &       &  km s$^{-1}$ &  km s$^{-1}$ &         &            \\
\multicolumn{1}{c}{(1)} & (2) & (3)  & (4)  &  (5)  &  (6)  &   (7)  & (8)        & \multicolumn{1}{c}{(9)}  &  \multicolumn{1}{c}{(10)}  \\
\hline
\multicolumn{6}{c}{Clusters of BM I type ($\Delta K\ge1.00$\,m)\,:} \\
A0038   & .1416 &   15 &   1.15 & $-$26.39 &  69~\,&  544 &  $-$99 & 2MASX J00281984+1354596     & 2MASX J00280539+1347225 \\
A0085A  & .0554 &  355 &   1.61 & $-$26.58 &  48*  & 1010 &     43 & MCG $-$02$-$02$-$086        & GIN 009                 \\
A0133A  & .0563 &  137 &   1.69 & $-$26.20 &  53*  &  760 &    204 & ESO 541$-$G013              & 2MASX J01013597$-$2203488 \\
A0150   & .0591 &   17 &   1.50 & $-$26.15 &  55~\,&  674 &    239 & UGC 00716                   & 2MASX J01085285+1320137 \\
A0152A  & .0594 &   88 &   1.67 & $-$26.04 &  39*  &  724 & $-$147 & 2MASX J01100320+1358417     & 2MASX J01100926+1407237 \\
A0193   & .0492 &   99 &   1.03 & $-$26.27 &  58~\,&  840 & $-$105 & CGCG 411$-$049              & CGCG 411$-$049            \\
A0208A  & .0796 &   66 &   1.12 & $-$26.28 &  34*  &  456 &  $-$27 & PGC 1169115                 & 2MASX J01303645+0027305 \\
A0225   & .0701 &    8 &   1.46 & $-$26.12 &  51~\,&      &        & 2MASX J01384892+1849311     & 2MASX J01384054+1848111 \\
A0261A  & .0473 &    9 &   1.69 & $-$25.71 &  57*  &      &        & 2MASX J01512719$-$0215317   & 2MASX J01520129$-$0210478 \\
A0279A  & .0800 &  101 &   1.33 & $-$26.44 &  59*  &  599 & $-$280 & MCG +00$-$06$-$002          & 2MASX J01570000+0103172 \\
A0376   & .0484 &  150 &   1.06 & $-$26.20 &  36~\,&  810 &    210 & UGC 02232                   & GIN 138                 \\
A0399   & .0720 &  101 &   1.92 & $-$26.83 &  57~\,& 1223 & $-$164 & UGC 02438                   & 2MASX J02580300+1251138 \\
A0401   & .0739 &  116 &   1.04 & $-$26.72 &  90~\,& 1144 &    173 & UGC 02450                   & 2MASX J02581064+1340419 \\
A0415   & .0810 &   14 &   1.89 & $-$26.57 &  67~\,&  617 & $-$671 & 2MASX J03065268$-$1206234   & 2MASX J03072144$-$1201357 \\
A0644   & .0693 &   44 &   1.13 & $-$26.41 &  42~\,&  700 &    103 & 2MASX J08172559$-$0730455   & 2MASX J08172714$-$0736025 \\
A0655   & .1272 &   61 &   1.58 & $-$27.24 & 142~\,&  729 &    472 & 2MASX J08252902+4707598     & 2MASX J08260055+4702348 \\
A0690A  & .0803 &   93 &   1.28 & $-$26.84 &  41*  &  540 & $-$405 & 2MASX J08391582+2850389     & 2MASX J08401564+2850447 \\
A0705A  & .1042 &   33 &   1.53 & $-$26.48 &  26*  &  615 &    165 & 2MASX J08474520+3001335     & 2MASX J08482745+2952036 \\
A0912A  & .0444 &   18 &   1.65 & $-$25.53 &  18*  &  356 &     38 & CGCG 008$-$008              & 2MASX J09590714+0000385 \\
A0941   & .1048 &   13 &   1.12 & $-$25.91 &  56~\,&  238 & $-$125 & 2MASX J10094349+0337229     & 2MASX J10094422+0337499 \\
A0971A  & .0929 &   48 &   1.28 & $-$26.55 &  41*  &  760 & $-$384 & 2MASX J10195207+4059179     & 2MASX J10194571+4059389 \\
A1004   & .1418 &   13 &   1.30 & $-$26.40 &  76~\,&  365 &     64 & 2MASX J10253527+5105541     & 2MASX J10254126+5106166 \\
A1023   & .1169 &    6 &   1.50 & $-$25.94 &  31~\,&      &        & LCRS B102528.0$-$063237     & LCRS B102529.5$-$063045   \\
A1068   & .1382 &   13 &   1.12 & $-$26.44 &  71~\,&  619 &    127 & 2MASX J10404446+3957117     & 2MASX J10403391+4003497 \\
A1076   & .1170 &   21 &   1.06 & $-$26.03 &  50~\,&  420 &    183 & 2MASX J10451352+5808334     & 2MASX J10453036+5812322 \\
A1146   & .1412 &   72 &   1.48 & $-$27.43 & 147~\,& 1019 & $-$324 & 2MASX J11011449$-$2243525   & 2MASX J11013225$-$2247390 \\
A1227A  & .1113 &   45 &   1.18 & $-$26.06 &  74*  &  733 &    150 & 2MASX J11213588+4802522     & 2MASX J11220532+4806152 \\
A1302   & .1156 &   58 &   1.70 & $-$26.56 &  85~\,&  767 &  $-$90 & 2MASX J11331462+6622454     & 2MASX J11305335+6630438 \\
A1308A  & .0501 &   56 &   1.09 & $-$26.02 &  25*  &  754 &    334 & PGC 035654                  & 2MASX J11325072$-$0347274 \\
A1413   & .1417 &   47 &   1.30 & $-$27.02 & 196~\,&  674 &    245 & 2MASX J11551798+2324177     & 2MASX J11551747+2323287 \\
A1516A  & .0769 &   72 &   1.09 & $-$26.30 &  45*  &  680 & $-$298 & 2MASX J12185235+0514443     & 2MASX J12185824+0515163 \\
A1644   & .0465 &  307 &   1.23 & $-$26.69 &  92~\,& 1030 &    150 & 2MASX J12571157$-$1724344   & 2MASX J12574919$-$1732431 \\
A1651   & .0841 &  222 &   1.28 & $-$26.45 &  70~\,&  960 &    190 & 2MASX J12592251$-$0411460   & 2MASX J12593749$-$0406597 \\
A1654   & .0840 &   25 &   1.23 & $-$26.32 &  31~\,&  512 &    141 & 2MASX J12592001+3001300     & 2MASX J12581976+2950557 \\
A1663A  & .0826 &  101 &   1.21 & $-$26.21 &  50*  &  705 &    453 & 2MASX J13025254$-$0230590   & 2MASX J13025000$-$0226380 \\
A1738   & .1173 &   59 &   1.50 & $-$26.93 &  82~\,&  546 & $-$381 & MCG +10$-$19$-$068          & 2MASX J13240096+5739160 \\
A1795   & .0628 &  179 &   1.04 & $-$26.33 & 115~\,&  835 &    254 & CGCG 162$-$010              & 2MASX J13482545+2624383 \\
A1809A  & .0793 &  132 &   1.01 & $-$26.44 &  74*  &  690 & $-$163 & 2MASX J13530637+0508586     & 2MASX J13523104+0456048 \\
A1837   & .0694 &   50 &   2.24 & $-$26.83 &  50~\,&  601 & $-$179 & 2MASX J14013635$-$1107431   & 2MASX J14012568$-$1109151 \\
A1864A  & .0867 &   61 &   1.15 & $-$26.39 &  68*  &  771 &    185 & 2MASX J14080526+0525030     & 2MASX J14070976+0520132 \\
A1890   & .0574 &   94 &   1.60 & $-$26.46 &  37~\,&  514 &    193 & NGC 5539                    & NGC 5535                \\
A1925   & .1064 &   55 &   1.18 & $-$26.34 &  92~\,&  718 &     33 & 2MASX J14283842+5651381     & 2MASX J14275634+5643558 \\
A2029   & .0775 &  202 &   1.85 & $-$27.25 &  82~\,& 1330 &    150 & IC 1101                     & 2MASX J15110004+0546578 \\
A2067A  & .0767 &  171 &   1.41 & $-$26.37 &  43*  &  850 & $-$658 & CGCG 165$-$049              & 2MASX J15234742+3111432 \\
A2107   & .0416 &  170 &   1.10 & $-$26.20 &  51~\,&  611 &    130 & UGC 09958                   & CGCG 136$-$050            \\
A2110   & .0981 &   53 &   1.24 & $-$25.94 &  54~\,&  472 & $-$105 & 2MASX J15395079+3043037     & 2MASX J15401322+3046338 \\
A2124   & .0667 &  118 &   2.00 & $-$26.45 &  50~\,&  787 &  $-$20 & UGC 10012                   & 2MASX J15444687+3557004 \\
A2128A  & .0583 &    5 &   1.46 & $-$26.06 &  30*  &      &        & 2MASX J15484313$-$0259344   & 2MASX J15474475$-$0252145 \\
A2170B  & .1052 &   33 &   1.09 & $-$25.80 &  21*  &  498 &  $-$45 & 2MASX J16165982+2311109     & 2MASX J16165866+2307549 \\
A2228   & .1005 &   30 &   1.66 & $-$26.53 &  55~\,&  794 &     53 & 2MASX J16474406+2956314     & 2MASX J16480084+2956575 \\
A2244   & .0997 &  106 &   1.81 & $-$26.81 &  89~\,& 1037 &      7 & 2MASX J17024247+3403363     & 2MASX J17021662+3358503 \\
A2271   & .0586 &   20 &   2.03 & $-$26.07 &  35~\,&  894 & $-$536 & CGCG 355$-$030              & 2MASX J17182094+7802142 \\
A2420   & .0852 &   10 &   1.11 & $-$26.69 &  88~\,&  712 & $-$454 & 2MASX J22101878$-$1210141   & 2MASX J22100145$-$1219291 \\
\hline
\end{tabular}
\end{minipage}
\end{table*}

\setcounter{table}{0}
\begin{table*}
\centering
\begin{minipage}{180mm}
\caption{ -- continued}
\begin{tabular}{@{}lcrrcrrrll@{}}
\hline
Abell & $z$ & $N_z$ & $\Delta K$ & $M_K$ & $N_A$ & $\sigma_v$ & $v_{pec}$ & \multicolumn{1}{c}{cD~galaxy ID} & \multicolumn{1}{c}{2nd-brightest~galaxy ID} \\
Cluster  &  &       &   mag     &  mag  &       &  km s$^{-1}$ &  km s$^{-1}$ &         &            \\
\multicolumn{1}{c}{(1)} & (2) & (3)  & (4)  &  (5)  &  (6)  &   (7)  & (8)        & \multicolumn{1}{c}{(9)}  &  \multicolumn{1}{c}{(10)}  \\
\hline
A2457   & .0589 &  113 &   1.00 & $-$26.70 &  53~\,&  620 & $-$250 & 2MASX J22354078+0129053     & 2MASX J22352797+0128153 \\
A2480   & .0725 &   12 &   1.34 & $-$26.20 & 108~\,&  806 &$-$1020 & 2MASX J22455898$-$1737320   & 2MASX J22460031$-$1741210 \\
A2544   & .0673 &   11 &   1.57 & $-$25.80 &  31~\,&  299 & $-$119 & 2MASX J23101507$-$1047540   & 2MASX J23095939$-$1102380 \\
A2589   & .0421 &   94 &   1.47 & $-$26.13 &  40~\,&  790 &   $-$9 & NGC 7647                    & 2MASX J23235357+1652479 \\
A2637   & .0712 &   11 &   1.22 & $-$26.19 &  60~\,&  579 &     33 & 2MASX J23385333+2127528     & 2MASX J23384222+2130038 \\
A2670   & .0766 &  256 &   1.08 & $-$26.65 & 142~\,&  881 &    430 & 2MASX J23541371$-$1025084   & 2MASX J23534052$-$1024201 \\
A2694   & .0974 &    8 &   2.37 & $-$26.72 & 132~\,&      &        & 2MASX J00022410+0823541     & 2MASX J00014206+0818145 \\
A2700A  & .0949 &   11 &   1.43 & $-$27.02 &  41*  &  780 &    450 & 2MASX J00034964+0203594     & 2MASX J00032769+0207014 \\
A3009   & .0652 &   23 &   1.87 & $-$26.38 &  54~\,&  447 &    115 & 2MASX J02220707$-$4833495   & FAIRALL 0379            \\
A3104   & .0727 &   89 &   1.15 & $-$25.85 &  37~\,&  750 &  $-$15 & LCRS B031238.4$-$453620     & LCRS B031257.9$-$453607   \\
A3109A  & .0631 &   10 &   1.05 & $-$26.03 &  15*  &  378 & $-$327 & 2MASX J03163934$-$4351169   & 2MASX J03160986$-$4333339 \\
A3112B  & .0751 &  112 &   1.39 & $-$26.83 &  95*  &  810 &    215 & ESO 248$-$G006              & LCRS B031515.1$-$442704   \\
A3120   & .0697 &    6 &   1.30 & $-$25.82 &  40~\,&      &        & 2MASX J03215645$-$5119357   & 2MASX J03223357$-$5127128 \\
A3407   & .0421 &   53 &   1.02 & $-$26.28 &  57~\,&  658 & $-$236 & ESO 207$-$G019              & 2MASX J07035803$-$4904502 \\
A3490   & .0687 &   88 &   1.52 & $-$26.35 &  91~\,&  680 & $-$187 & 2MASX J11452010$-$3425596   & 2MASX J11453744$-$3420143 \\
A3571   & .0385 &  172 &   1.00 & $-$26.63 & 126~\,&  880 & $-$190 & ESO 383$-$G076              & 2MASX J13485033$-$3309071 \\
A3854A  & .1231 &   23 &   1.25 & $-$26.62 &  60*  &  492 &  $-$24 & 2MASX J22174585$-$3543293   & 2MASX J22182696$-$3526418 \\
A4059   & .0488 &  188 &   1.29 & $-$26.61 &  66~\,&  718 &    440 & ESO 349$-$G010              & MCG $-$06$-$01$-$006          \\[.5ex]
\multicolumn{7}{c}{Clusters of intermediate type ($0.70\,m<\Delta K<1.00\,m$)\,:} \\
A0126   & .0548 &   11 &   0.71 & $-$25.79 &  51~\,&  530 & $-$490 & 6dF J0059591$-$135943       & 2MASX J00595379$-$1414403 \\
A0478   & .0862 &   13 &   0.81 & $-$26.45 & 104~\,&  944 &  $-$86 & 2MASX J04132526+1027551     & 2MASX J04125893+1035156 \\
A0715   & .1432 &   17 &   0.97 & $-$26.27 &  69~\,&  994 &  $-$73 & 2MASX J08545745+3524513     & 2MASX J08543921+3522043 \\
A1406B  & .1175 &   15 &   0.92 & $-$25.97 &  40*  &  332 &    143 & 2MASX J11530531+6753513     & 2MASX J11530926+6748103 \\
A1668   & .0638 &   95 &   0.83 & $-$25.86 &  54~\,&  759 & $-$113 & IC 4130                     & IC 4139                 \\
A1749A  & .0561 &   80 &   0.94 & $-$26.06 &  45*  &  451 &  $-$28 & IC 4269                     & IC 4271 NED01           \\
A1767   & .0713 &  159 &   0.79 & $-$26.60 &  65~\,&  863 &     70 & MCG +10$-$19$-$096          & 2MASX J13342636+5922256 \\
A2148   & .0885 &   47 &   0.75 & $-$26.01 &  41~\,&  489 &    425 & GIN 478                     & GIN 484                 \\
A2372   & .0600 &    7 &   0.76 & $-$25.63 &  42~\,&      &        & 2MASX J21451552$-$1959406   & 2MASX J21452542$-$2007300   \\
A2401   & .0576 &   35 &   0.91 & $-$26.07 &  66~\,&  438 &    142 & 2MASX J21582246$-$2006145   & PKS 2156$-$203            \\
A2593A  & .0424 &  121 &   0.83 & $-$26.07 &  40*  &  644 &    110 & NGC 7649                    & CGCG 431$-$056            \\
A2622   & .0620 &   57 &   0.99 & $-$25.95 &  41~\,&  942 & $-$171 & PGC 071807                  & 2MASX J23352311+2719220 \\
A2626A  & .0585 &   96 &   0.96 & $-$26.36 &  43*  & 1057 & $-$802 & IC 5338                     & IC 5337                 \\
A2734   & .0612 &  189 &   0.71 & $-$26.14 &  58~\,&  879 &    200 & ESO 409$-$G025              & SARS 002.29825$-$29.21075 \\
A2871B  & .1215 &   53 &   0.76 & $-$25.59 &  60*  &  319 &  $-$50 & 2MASX J01075037$-$3643217   & SARS 016.31773$-$36.88759 \\
A2961A  & .1246 &   24 &   0.99 & $-$25.99 &  20*  &  539 &  $-$33 & 2MASX J02000056$-$3114133   & 2MASX J02000195$-$3119133 \\
A2984   & .1038 &   33 &   0.86 & $-$26.05 &  54~\,&  571 &    318 & ESO 298$-$G017              & 2MASX J02105846$-$4011169 \\
A3301   & .0534 &   38 &   0.82 & $-$26.11 & 172~\,&  686 &     34 & NGC 1759                    & 2MASX J05013224$-$3844105 \\
A3376   & .0453 &  165 &   0.73 & $-$25.90 &  42~\,&  831 &  $-$14 & ESO 307$-$G013              & 2MASX J06020973$-$3956597 \\
A3556   & .0473 &  209 &   0.80 & $-$26.28 &  49~\,&  698 &     22 & ESO 444$-$G025              & 2MASX J13235763$-$3138453 \\
A3558   & .0474 &  509 &   0.88 & $-$26.88 & 226~\,&  940 & $-$302 & ESO 444$-$G046              & 2MASX J13272961$-$3123237 \\
A3998   & .0899 &   17 &   0.72 & $-$26.02 &  40~\,&  574 &     90 & ESO 347$-$G009              & LCRS B231932.9$-$421633   \\[.5ex]
\multicolumn{5}{c}{Clusters of NBMI type ($\Delta K\le0.70\,m$)\,:} \\
A0076   & .0407 &   13 &   0.34 & $-$26.08 &  42~\,&  459 & $-$677 & IC 1565                     & IC 1568                 \\
A0119   & .0447 &  339 &   0.53 & $-$26.44 &  69~\,&  840 &     25 & UGC 00579                   & UGC 583                 \\
A0367   & .0899 &   33 &   0.38 & $-$25.81 & 101~\,&  900 &    539 & 2MASX J02363713$-$1922168   & 2MASX J02362667$-$1915078 \\
A0389   & .1131 &   55 &   0.15 & $-$26.41 & 133~\,&  759 &     41 & 2MASX J02512479$-$2456393   & 2MASX J02513267$-$2504233 \\
A0754   & .0538 &  470 &   0.57 & $-$26.26 &  92~\,&  976 &     59 & 2MASX J09083238$-$0937470   & 2MASX J09101737$-$0937068 \\
A1149   & .0714 &   49 &   0.15 & $-$25.45 &  34~\,&  313 & $-$292 & 2MASX J11025750+0736136     & 2MASX J11032040+0730463 \\
A1168   & .0908 &   46 &   0.34 & $-$25.93 &  52~\,&  597 &    307 & 2MASX J11071768+1551475     & 2MASX J11080366+1554133 \\
A1222   & .1120 &   45 &   0.43 & $-$26.24 &  75~\,&  523 & $-$143 & 2MASX J11201257+4711323     & MCG +08$-$21$-$017          \\
A1361   & .1154 &   20 &   0.70 & $-$26.02 &  57~\,&  456 &    204 & 2MASX J11433959+4621202     & 2MASX J11424122+4624363 \\
A1630A  & .0649 &   37 &   0.44 & $-$25.78 &  41*  &  440 & $-$216 & CGCG 043$-$047 NED01        & CGCG 043$-$044            \\
A1650   & .0836 &  220 &   0.40 & $-$25.80 & 114~\,&  789 &    117 & 2MASX J12584149$-$0145410   & 2MASX J12583829$-$0134290 \\
A1691   & .0722 &  111 &   0.55 & $-$26.51 &  64~\,&  843 &     90 & MCG +07$-$27$-$039          & 2MASX J13100997+3909235 \\
A1736B  & .0448 &  148 &$-$0.69 & $-$25.71 &  68*  &  860 & $-$148 & ESO 509$-$G009              & IC 4252                 \\
A1800   & .0755 &   91 &   0.67 & $-$26.57 &  40~\,&  723 &     28 & UGC 08738                   & 2MASX J13493157+2800016 \\
A1814   & .1262 &   39 &   0.67 & $-$26.15 &  71~\,&  590 &     98 & 2MASX J13540294+1454409     & 2MASX J13534984+1445380 \\
A1839   & .1295 &   49 &   0.28 & $-$25.52 &  63~\,& 1104 &    316 & 2MASX J14023276$-$0451249   & 2MASX J14023417$-$0449449 \\
A1918B  & .1408 &   23 &   0.48 & $-$26.37 & 105*  &  825 & $-$511 & 2MASX J14252238+6311524     & 2MASX J14252117+6309214 \\
A1920   & .1314 &   39 &   0.31 & $-$25.90 & 103~\,&  562 & $-$120 & 2MASX J14272450+5545009     & 2MASX J14270303+5553259 \\
\hline
\end{tabular}
\end{minipage}
\end{table*}

\setcounter{table}{0}
\begin{table*}
\centering
\begin{minipage}{180mm}
\caption{ -- continued}
\begin{tabular}{@{}lcrrcrrrll@{}}
\hline
Abell & $z$ & $N_z$ & $\Delta K$ & $M_K$ & $N_A$ & $\sigma_v$ & $v_{pec}$ & \multicolumn{1}{c}{cD~galaxy ID} & \multicolumn{1}{c}{2nd-brightest~galaxy ID} \\
Cluster  &  &       &   mag     &  mag  &       &  km s$^{-1}$ &  km s$^{-1}$ &         &            \\
\multicolumn{1}{c}{(1)} & (2) & (3)  & (4)  &  (5)  &  (6)  &   (7)  & (8)        & \multicolumn{1}{c}{(9)}  &  \multicolumn{1}{c}{(10)}  \\
\hline
A1927   & .0949 &   50 &   0.58 & $-$25.89 &  50~\,&  650 &    376 & 2MASX J14310681+2538013     & 2MASX J14303458+2538495 \\
A1991   & .0589 &  135 &   0.69 & $-$26.03 &  60~\,&  625 &    320 & NGC 5778                    & CGCG 105$-$068            \\
A2050   & .1190 &   37 &   0.70 & $-$25.76 &  50~\,&  688 & $-$209 & 2MASX J15161794+0005203     & 2MASX J15160950+0014541 \\
A2051   & .1180 &   54 &$-$0.31 & $-$25.90 &  94~\,&  535 &    170 & 2MASX J15164416$-$0058096   & 2MASX J15165808$-$0106394 \\
A2079A  & .0667 &  151 &   0.20 & $-$26.44 &  49*  &  816 & $-$318 & UGC 09861 NED02             & UGC 09861 NED01         \\
A2089   & .0731 &  105 &   0.69 & $-$26.00 &  70~\,&  722 &    209 & 2MASX J15324982+2802224     & 2MASX J15325912+2753405 \\
A2147   & .0365 &  397 &   0.20 & $-$25.51 &  52~\,&  890 & $-$203 & UGC 10143                   & UGC 10143 NOTES02       \\
A2428   & .0845 &   51 &   0.47 & $-$26.23 &  51~\,&  453 &    173 & 2MASX J22161561$-$0919590   & 2MASX J22164131$-$0914138 \\
A2554   & .1109 &   89 &   0.42 & $-$25.99 & 100~\,&  717 & $-$505 & 2MASX J23121995$-$2130098   & 2MASX J23121357$-$2130018 \\
A2572   & .0388 &  107 &   0.66 & $-$25.62 &  32~\,&  620 & $-$290 & NGC 7571                    & NGC 7598                \\
A2597   & .0830 &   45 &   0.47 & $-$25.40 &  43~\,&  564 & $-$200 & PGC 071390                  & 2MASX J23245745$-$1212001 \\
A2657   & .0409 &   64 &   0.24 & $-$25.37 &  51~\,&  782 &      0 & CGCG 407$-$053 NED02        & CGCG 407$-$050            \\
A2969   & .1252 &   20 &$-$0.15 & $-$26.17 &  83~\,&  850 &    411 & 2MASX J02033533$-$4106002   & LCRS B020108.7$-$412348   \\
A3093   & .0828 &   26 &   0.64 & $-$25.95 &  93~\,&  419 &  $-$99 & AM 0309$-$473 NED02         & AM 0309$-$473 NED04       \\
A3144   & .0444 &   31 &   0.44 & $-$25.48 &  54~\,&  532 & $-$507 & 2MASX J03370557$-$5501186   & IC 1987 NED02           \\
A3546   & .1065 &   14 &   0.36 & $-$26.15 &  39~\,&  275 & $-$132 & 2MASX J13130596$-$2958432   & 2MASX J13140646$-$3011328 \\
A3562   & .0471 &  265 &   0.33 & $-$25.60 & 129~\,& 1070 &    241 & ESO 444$-$G072              & 2MASX J13350306$-$3139187 \\
\hline
\end{tabular}
\end{minipage}
\end{table*}
 
 

\begin{thebibliography}{}
\setlength{\labelwidth}{0pt}

\bibitem[\protect\citeauthoryear{Abell et al.}{1989}]{b1} Abell G.O., Corwin Jr., H.G., Olowin R.P. 1989, ApJS, 70, 1

\bibitem[]{b2} Andernach H., Tago E., Einasto M., Einasto J., Jaaniste J.,
2005, Nearby Large-Scale Structures and the Zone of Avoidance,
eds. A.P.~Fairall \& P.~Woudt, ASP Conf.\ Series 329, 283

\bibitem[]{b3} Andreon S., Garilli B., Maccagni D., Gregorini L., Vettolani G., 1992, A\&A, 266, 127

\bibitem[]{b4} Arag\'on-Salamanca A., Baugh C.M., Kauffmann G., 1998, MNRAS, 297, 427

\bibitem[]{b5} Baier F.W., Schmidt K.-H., AN, 1992, 313, 275

\bibitem[]{b6} Barnes J.E., 1989, Nature 338, 123

\bibitem[]{b7} Bautz L., Morgan W.W., 1970, ApJ, 162, L149

\bibitem[]{b8} Bell E.F., McIntosh D.H., Katz N., Weinberg M.D. 2003, ApJS, 149, 289

\bibitem[]{bell} Bell E.F., et al.\ 2006, ApJ, 640, 241

\bibitem[]{bern} Bernardi M., Hyde J.B., Sheth R.K., Miller C.J., Nichol R.C. 2007, AJ, 133, 1741

\bibitem[]{b9} Binggeli B., 1982, A\&A, 107, 338

\bibitem[]{cdb10} Courteau S., Dutton A.A., van\,den\,Bosch F.C., MacArthur L.A.,
       Dekel A., McIntosh D.H., Dale D.A. 2007, ApJ, 671, 203

\bibitem[]{b11} Cowie L.L., Binney J., 1977, ApJ, 215, 723

\bibitem[]{b12} Coziol R., Andernach H., Caretta C.A., Alamo Mart\'{\i}nez K.A., Tago E., 2009, AJ, 137, 4795

\bibitem[]{del} De\,Lucia G., Springel V., White S.D.M., Croton D., Kauffmann G., 2006, MNRAS, 366, 499

\bibitem[]{b13} De\,Lucia G., Blaizot J., 2007, MNRAS, 375, 2

\bibitem[]{b14} Dressler A., 1980, ApJ, 236, 351

\bibitem[]{b15} Dubinski J., 1998, ApJ, 502, 141

\bibitem[]{b16} Fabian A.C., 1994, ARA\&A, 32, 277

\bibitem[]{b17} Forman W, Jones C. 1982, ARA\&A, 20, 547

\bibitem[]{b18} Fuller T.M., West M.J., Bridges T.J., 1999, ApJ, 519, 22

\bibitem[]{b20} Gallagher J.S., Ostriker J.P., 1972, AJ, 77, 288

\bibitem[]{b19} Gao L., Loeb A., Peebles P.J.E., White S.D.M., Jenkins A., 2004, ApJ, 614, 17

\bibitem[]{b21} Garijo A., Athanassoula E., Garcia-G\'{o}mez C., 1997, A\&A, 327, 930

\bibitem[]{b22} Hansen S.M., Sheldon E.S., Wechsler R.H., Koester B.P., 2009, ApJ, 699, 1333

\bibitem[]{b23} Hausman M.A., Ostriker J.P., 1978, ApJ, 224, 320

\bibitem[]{b24} Hill J.M., Hintzen P., Oegerle W.R., Romanishin W., Lesser M.P.,
Eisenhamer J.D., Batuski D.J. 1988, ApJ, 332, L23

\bibitem[]{b24a} Hudson D.S., Mittal R., Reiprich T.H., Nulsen P.E.J., Andernach H., 
Sarazin C.L., 2010, A\&A 513, A37

\bibitem[]{b25} Jarrett T.H., Chester T., Cutri R, Schneider S., Skrutskie M.,
Huchra J.P., 2000, AJ, 119, 2498

\bibitem[]{b26} Jord\'an A., C\^ot\'e P., West M.J., Marzke R.O., Minniti D., Rejkuba M. 2004, AJ, 127 24

\bibitem[]{b27} Kaastra J.S., Ferrigno C., Tamura T., Paerels F.B.S.,
Peterson J.R., Mittaz J.P.D. 2001, A\&A, 365, L99

\bibitem[]{b28} Knebe A., Gill S.P.D., Gibson B.K., Lewis G.F.,
Ibata R.A., Dopita M.A., 2004, ApJ, 603, 7

\bibitem[]{b29} Kochanek C.S., Pahre M.A., Falco E.E. et al., 2001, ApJ, 560, 566

\bibitem[]{b30} Kormendy J., Djorgovski S., 1989, ARA\&A, 27, 235

\bibitem[]{b31} Lambas D., Groth E.J., Peebles P.J.E., 1988, AJ, 95, 996

\bibitem[]{b32} Lauer T.R., Faber S.M., Richstone D. et al., 2007, ApJ, 662, 808

\bibitem[]{b34} Lin Y.-T., Mohr J.J., 2004, ApJ, 617, 879

\bibitem[]{mr} Malumuth E.M., Richstone D.O., 1984, ApJ, 276, 413


\bibitem[]{b35} Masters K.L., Springob C.M., Huchra J.P., 2008, AJ, 135, 1738

\bibitem[]{b36} Matthews T.A., Morgan W.W., Schmidt M. 1964, ApJ, 140, 35

\bibitem[]{b37} Merritt D., 1983, ApJ, 264, 24

\bibitem[]{b38} Merritt D., 1984, ApJ, 276, 26

\bibitem[]{b39} Merritt D., 1985, ApJ, 289, 18

\bibitem[]{b40} Niederste-Ostholt M., Strauss M.A., Dong F., Koester B.P., McKay T.A., 2010, MNRAS, 405, 2023

\bibitem[]{b41} Oegerle W.R., Hill J.M. 1994, AJ, 107, 857

\bibitem[]{b42} Oegerle W.R., Hill J.M. 2001, AJ, 122, 2858

\bibitem[]{b43} Ostriker J.P., Tremaine S.D., 1975, ApJ, 202, L113

\bibitem[]{b44} Ostriker J.P., Hausman M.A. 1977, ApJ, 217, L125

\bibitem[]{b45} Peterson J.R., Paerels F.B.S., Kaastra J.S. et al.\ 2001, A\&A, 365, L104

\bibitem[]{b46} Piffaretti R., Arnaud M., Pratt G.W., Pointecouteau E., Melin J.-B., 2011, A\&A, 534, A109

\bibitem[]{b47} Pimbblet K.A., Roseboom I.G., Doyle M.T., 2006, MNRAS, 368, 651


\bibitem[]{b48b} Postman M., Huchra J.P., Geller M.J., 1985, AJ, 90, 1400

\bibitem[]{b82} Quintana H., Lawrie D.C., 1982, 97, 1


\bibitem[]{b50} Rhee G., Katgert P., 1987, A\&A, 183, 217

\bibitem[]{b51} Richstone D.O., 1975, ApJ, 200, 535

\bibitem[]{b52} Richstone D.O., 1976, ApJ, 204, 642

\bibitem[]{b53} Searle L., Sargent W.L.W., Bagnuolo W.G., 1973, ApJ, 179, 427

\bibitem[]{b54} Schlegel D.J., Finkbeiner D.P., Davis M., 1998, ApJ, 500, 525

\bibitem[]{b55} Schneider D., Gunn J.E., Hoessel J.G., 1983, ApJ, 268, 476

\bibitem[]{b56} Schombert J.M., 1987, ApJS, 64, 643

\bibitem[]{b57} Schombert J.M., 1988, ApJ, 328, 475

\bibitem[]{b58} Schombert J.M., 1992, in {\it Morphological and Physical
Classification of Galaxies}, Dordrecht: Kluwer Academic Publishers,
eds.\ G.~Longo, M.~Capaccioli, G.~Busarello, Astrophysics and
Space Science Library, 178, p.\ 53 

\bibitem[]{b59a} Scott E.L., 1957, AJ 62, 248

\bibitem[]{b59} Sharples R.M., Ellis R.S., Gray P.M., 1988, MNRAS, 231, 479

\bibitem[]{b60} Silk J., 1976, ApJ, 208, 646


\bibitem[]{b62} Struble M.F., 1987, ApJ, 317, 688

\bibitem[]{b63} Tamura T., Kaastra J.S., Peterson J.R. et al.\ 2001, A\&A, 365, L87

\bibitem[]{b64} Temi P., Brighenti F., Mathews W.G., 2008, ApJ, 672, 244

\bibitem[]{b65} Tonry J.L. 1987, Structure and Dynamics of Elliptical
Galaxies, ed.~T.~de\,Zeeuw, IAU Symp.\ 127, 89, Reidel, Dordrecht

\bibitem[]{b66} Torlina L., De\,Propris R.D., West M.J., 2007, ApJ, 660, L97

\bibitem[]{b66} Tovmassian H.M., 2010, RMxA\&A, 46, 61

\bibitem[]{tpa04} Tovmassian H.M., Plionis M., Andernach H., 2004, ApJ, 617, L111

\bibitem[]{tdl} Tran K.-V. H., van~Dokkum P., Illingworth G.D., Kelson D., Gonzalez A., Franx M., 2005, ApJ, 619, 134

\bibitem[]{b67} Tremaine S., 1990, Dynamics and Interactions of Galaxies, 
    ed.\ R.~Wielen, p.\ 394, Springer Verlag, Berlin

\bibitem[]{b68} Tutukov A.V., Dryumov V.V., Dryumova G.N.,
2007, Astronomy Reports 51, 435

\bibitem[]{b69} van~Dokkum P.G., 2005, AJ, 130, 2647


\bibitem[]{b70} von der Linden A., Best P.N., Kauffmann G., White S.D.M., 2007, MNRAS, 379, 867

\bibitem[]{b71} West M.J., Jones C., Forman W., 1995, ApJ, 451, L5

\bibitem[]{b72} Whiley I.M., Arag\'on-Salamanca A, De Lucia G. et al.\ 2008, MNRAS, 387, 1253

\bibitem[]{b73} White S.D.M., 1976, MNRAS, 174, 19



\bibitem[]{b75} Zabludoff A.L., Mulchaey J.S. 1998, ApJ, 496, 39

\bibitem[]{b76} Zhang Y.-Y., Andernach H., Caretta C., Reiprich T.H., B\"{o}hringer H.,
    Puchwein E., Sijacki D., Girardi M., 2011, A\&A, 526, A105

\end{thebibliography}
\end{document}